\documentclass{chi2010}
\usepackage{times}
\usepackage{url}
\usepackage{graphics}
\usepackage{color}
\usepackage[pdftex]{hyperref}
\hypersetup{%
pdftitle={Your Title}, pdfauthor={Your Authors}, pdfkeywords={your
keywords}, bookmarksnumbered, pdfstartview={FitH}, colorlinks,
citecolor=black, filecolor=black, linkcolor=black, urlcolor=black,
breaklinks=true, }

\newcommand{\comment}[1]{}
\definecolor{Orange}{rgb}{1,0.5,0}

\newcommand{\xhdr}[1]{{\bf #1}}
\newcommand{\omt}[1]{}

\enlargethispage{\baselineskip} 

\newcommand{\denselist}{ \itemsep -5pt\topsep-10pt\partopsep-10pt }
\newcommand{\denselistA}{ \itemsep -2pt\topsep-5pt\partopsep-7pt }

\begin{document}

\setlength{\paperheight}{11in}
\setlength{\paperwidth}{8.5in}
\setlength{\pdfpageheight}{\paperheight}
\setlength{\pdfpagewidth}{\paperwidth}

\title{
Signed Networks in Social Media
} \numberofauthors{3}
\author{
\alignauthor Jure Leskovec \\
\affaddr{Stanford University}\\
\email{jure@cs.stanford.edu}
\alignauthor Daniel Huttenlocher \\
\affaddr{Cornell University}\\
\email{dph@cs.cornell.edu}
\alignauthor Jon Kleinberg \\
\affaddr{Cornell University}\\
\email{kleinber@cs.cornell.edu}
}

\maketitle

\begin{abstract}
Relations between users on social media sites often reflect a mixture of
positive (friendly) and negative (antagonistic) interactions. In contrast to
the bulk of research on social networks that has focused almost exclusively on
positive interpretations of links between people, we study how the interplay
between positive and negative relationships affects the structure of on-line
social networks. We connect our analyses to theories of signed networks from
social psychology. We find that the classical theory of structural balance
tends to capture certain common patterns of interaction, but that it is also at
odds with some of the fundamental phenomena we observe --- particularly related
to the evolving, directed nature of these on-line networks.  
We then develop an alternate theory of status that better
explains the observed edge signs and provides insights into the
underlying social mechanisms. Our work provides one of the first
large-scale evaluations of theories of signed networks using on-line
datasets, as well as providing a perspective for reasoning about
social media sites.
\end{abstract}

\vspace{-3mm}
.
\vspace{-3mm}

\section{Introduction}
\label{sec:intro}
Social network analysis provides a useful perspective on a range of social
computing applications. The structure of networks arising in such applications
offers insights into patterns of interactions, and reveals global phenomena at
scales that may be hard to identify when looking at a finer-grained resolution.
At the same time, there is an ongoing challenge in adapting such network
approaches to the study of social computing: users develop rich relationships
with one another in these settings, while network analyses generally reduce
these complex relationship to the existence of simple pairwise links. It is a
fundamental research problem to bridge the gap between the richness of the
existing relationships and the stylized nature of network representations of
these relationships.

The main focus of our work here is to examine the interplay between positive
and negative links in social media --- a dimension of on-line social 
network analysis that has been largely unexplored.
With relatively few exceptions (e.g.,
\cite{brzozowski-friend-foe,kunegis-slashdot-zoo,lampe-slashdot}), research in
on-line social networks has focused on contexts in which the interactions have
largely only positive interpretations --- that is, connecting people to their
friends, fans, followers, and collaborators.
But in many settings it is
important to also explicitly take negative relations into consideration,
especially when studying interactions in social media: discussion lists are
filled with controversy and disagreement, and social-networking sites harbor
antagonism alongside amity. The richness of a social network in such cases
generally consists of a mixture of both positive and negative interactions,
co-existing in a single structure.

We aim to develop a better understanding of the role that network structure
plays when some links between people are positive while others are negative.
For instance, in on-line rating sites such as Epinions, people can give both
positive and negative ratings not only to items but also to other raters.  In
on-line discussion sites such as Slashdot, users can tag other users as
``friends'' and ``foes''. Our approach here is to adapt and extend theories
from social psychology to analyze these types of signed networks 
as they arise in social computing applications. 
These theories enable us to characterize the differences between
the observed and predicted configurations of positive and negative links in
on-line social networks. We also use contrasts between the theories
to draw inferences about how links are being used in particular social
computing applications. In addition to insights into the applications
themselves, our studies provide, to the best of our knowledge, some of the
first large-scale evaluations of these social-psychological theories via
on-line datasets.

\xhdr{Positive and negative links in on-line data.} To carry out such an
investigation, we need two fundamental ingredients: (i) large-scale datasets
from social applications where the {\em sign} of each link --- whether it is
positive or negative --- can be reliably determined, and (ii) theories of
signed networks that help us reason about how different patterns of positive
and negative links provide evidence for the expression of different kinds of
relationships across these applications.

We investigate social network structures from three widely-used Web sites.
The first is the trust network of Epinions, where users create signed directed
relations to each other indicating trust or distrust. The second is the social
network of the technology blog Slashdot, where users designate others as
``friends'' or ``foes.''  The third is the network defined by votes for
Wikipedia admin candidates.  When a Wikipedia user is considered for a
promotion to the status of an admin, the community is able to cast public votes
in favor of or against the promotion of this admin candidate. We view a
positive vote as corresponding to a positive link from the voter to the
candidate, and a negative vote as a negative link. The Epinions and Slashdot
networks are explicitly presented to users as social networking features of the
sites, whereas in the case of Wikipedia the network interpretation is implicit.

The meanings of positive and negative signs are different across these
settings, and this is precisely the point: we wish to use theories of signed
edges to evaluate how the positive and negative edges are being used in each
setting, and to identify commonalities and differences in the underlying
networks in relatively different application contexts. Moreover, while the
current work focuses on domains in which the signs of edges are overtly denoted
(either explicitly by direct linking, or implicitly through actions such as
voting on Wikipedia), we believe the underlying issues reach more broadly into
any application where positive and negative attitudes between users can be
conveyed, such as through sentiment in text \cite{pang-lee-sentiment-book}.

\begin{figure}[t]
  \centering
  \small
  \begin{tabular}{cccc}
   \includegraphics[width=0.1\textwidth]{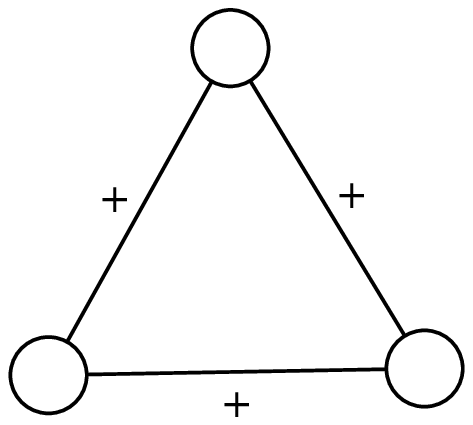} &
   \includegraphics[width=0.1\textwidth]{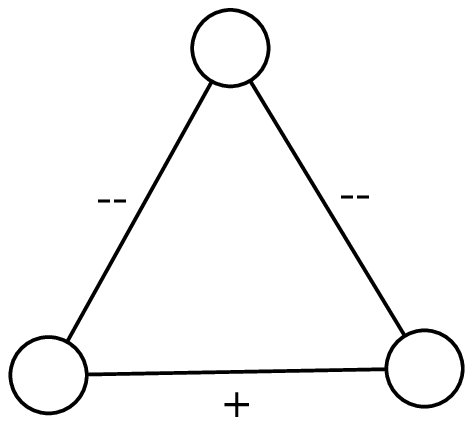} &
   \includegraphics[width=0.1\textwidth]{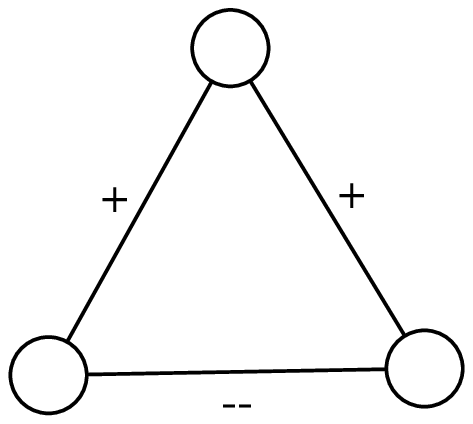} &
   \includegraphics[width=0.1\textwidth]{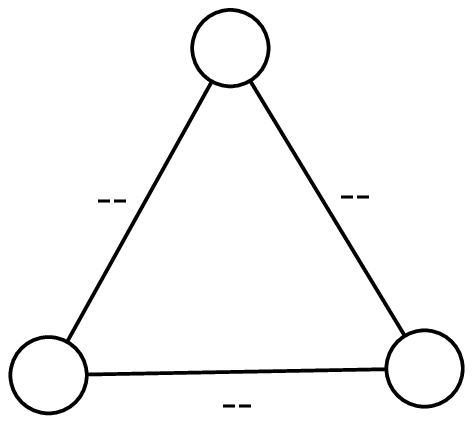} \\
   triad $T_3$ & triad $T_1$ & triad $T_2$ & triad $T_0$
\end{tabular}
    \vspace{-3mm}
    \caption{Undirected signed triads. Based on the
    number of positive edges we label triads with odd number of pluses
    as \emph{balanced} ($T_3,T_1$), and triads with even positive edges
    ($T_2, T_0$) as \emph{unbalanced}.}
    \vspace{-5mm}
    \label{fig:triads}
\end{figure}

\xhdr{Theories of signed networks: Balance.} We analyze these on-line signed
networks using two different theories, and a central issue in our study is the
extent to which each of these theories provides a plausible explanation for the
structure and dynamics of the observed networks.

The first of these theories is {\em structural balance theory}, which
originated in social psychology in the mid-20th-century. As formulated by
Heider in the 1940s \cite{heider-balance46}, and subsequently cast in
graph-theoretic language {by Cartwright and Harary}
\cite{cartwright-harary-balance}, structural balance considers the possible
ways in which triangles on three individuals can be signed, and posits that
triangles with three positive signs (three mutual friends,
Figure~\ref{fig:triads} $T_3$) and those with one positive sign (two friends
with a common enemy, Fig.~\ref{fig:triads} $T_1$) are more plausible --- and
hence should be more prevalent in real networks --- than triangles with two
positive signs (two enemies with a common friend, $T_2$) or none
(three mutual enemies, $T_0$). Balanced triangles with three positive edges
exemplify the principle that ``the friend of my friend is my friend,'' whereas
those with one positive and two negative edges capture the notions that ``the
friend of my enemy is my enemy,'' ``the enemy of my friend is my enemy,'' and
``the enemy of my enemy is my friend.''

Structural balance theory has been developed extensively in the time since this
initial work \cite{wasserman-faust}, including the formulation of a variant ---
{\em weak structural balance} --- proposed by Davis in the 1960s as a way of
eliminating the assumption that ``the enemy of my enemy is my friend''
\cite{davis-gen-balance}. In particular, weak structural balance posits that
only triangles with exactly two positive edges are implausible in real
networks, and that all other kinds of triangles should be permissible.

\xhdr{Theories of signed networks: Status.} Balance theory can be viewed as a
model of likes and dislikes. However, as Guha et al. observe in the context of
Epinions \cite{guha-trust}, a signed link from $A$ to $B$ can have more than
one possible interpretation, depending on $A$'s intention in creating the link.
In particular, a positive link from $A$ may mean, ``$B$ is my friend,'' but it
also may mean, ``I think $B$ has higher status than I do.'' Similarly, a
negative link from $A$ to $B$ may mean ``$B$ is my enemy'' or ``I think $B$ has
lower status than I do.''

Here we develop this idea into a new theory of {\em status}, which provides a
different organizing principle for directed networks 
of signed links. In this theory of
status, we consider a positive directed link to indicate that the creator of
the link views the recipient as having higher status; and a
negative directed link indicates that the recipient is viewed as 
having lower status.
These relative levels of status can then be propagated along multi-step paths
of signed links, often leading to different predictions than balance theory.

\xhdr{Comparing the two theories.} To give a sense for how the differences
between status and balance arise, consider the situation in which a user $A$
links positively to a user $B$, and $B$ in turn links positively to a user $C$.
If $C$ then forms a link to $A$, what sign should we expect this link to have?
Balance theory predicts that since $C$ is a friend of $A$'s friend, we should
see a positive link from $C$ to $A$. Status theory, on the other hand, predicts
that $A$ regards $B$ as having higher status, and $B$ regards $C$ as having
higher status --- so $C$ should regard $A$ as having low status and hence be
inclined to link negatively to $A$. In other words, the two theories suggest
opposite conclusions in this case.

Thus balance theory predicts that certain types of triads such as all-positive
cycles should be overrepresented compared to chance, whereas status theory
makes predictions that often differ.  We study all the possible types of signed
triads and the predictions made by the different theories.  In doing so we
consider several experimental conditions, including both directed and
undirected networks, as well as both respecting and ignoring the order in which
edges were created. For each such experimental condition we consider whether
the observed number of triads of each type is overrepresented or
underrepresented compared to chance, and contrast that with the predictions
made by the balance and status theories.  This analysis give us a picture of
the aggregate patterns of links in the social networks, and the degree to which
they are explained in terms of each theory.

\xhdr{Summary of Findings: Comparison of Balance and Status.} Both of these
theories concern relationships between people; by adapting them to our on-line
network datasets, they provide potentially informative perspectives on the link
structures we find there.

Balance theory was initially intended as a model for undirected networks,
although it has been commonly applied to directed networks by simply
disregarding the directions of the links \cite{wasserman-faust}. When we do
this, we find significant alignment between the observed network data and
Davis's notion of weak structural balance: triangles with exactly two positive
edges are massively underrepresented in the data relative to chance, while
triangles with three positive edges are massively overrepresented. In two of
the three datasets, triangles with three negative edges are also
overrepresented, which is at odds with Heider's formulation of balance theory.
These findings are already intriguing, since it has traditionally been
difficult to evaluate the predictions of structural balance theory on large
network datasets.  Rather, empirical investigations to date have generally
focused on small networks where social relations can be observed through direct
interaction with the individuals involved (see e.g.
\cite{doreian-balance-partitioning}). The trouble with assessing structural
balance at small scales is that one expects its predictions to be aggregate
rather than absolute --- that is, one expects to see certain kinds of triangles
as statistically more abundant or less abundant in the data, and the
significance of such biases towards certain kinds of triangles can stand out
much more clearly when they are accumulated over a large amount of data.

Ultimately, however, we would like to understand the networks in these on-line
systems as directed structures that evolve over time. When we view the network
data in this way, our main conclusion is that the theory of status is more
effective at explaining local patterns of signed links, and that it naturally
extends to capture richer aspects of user behavior, including heterogeneity in
their linking tendencies. For example in the case offered as an illustration
above, where user $A$ links positively to user $B$ and user $B$ links
positively to user $C$, we find that negative links from $C$ to $A$ are
massively overrepresented relative to chance, with positive links
correspondingly underrepresented.

\xhdr{Implications.} There are several potentially interesting implications of
our results. First, the comparison of balance and status provides insights into
ways in which people use linking mechanisms in social computing applications.
In particular, there are important domains such as rating reviewers on Epinions
and voting for admins on Wikipedia in which such links appear, in aggregate, to
be used more dominantly for expressions of status than for expressions of likes
and dislikes.

The contrast between balance and status is also related to the distinction
between undirected and directed interpretations of links. Our findings suggest
that it is important to understand the roles of different theories in both
undirected and directed representations of networks. Indeed, the theory of
status only makes sense with directed links --- since it posits a status
differential from the creator of a link to its recipient --- while the theory
of balance has been applied in both undirected and directed settings (e.g.,
\cite{wasserman-faust}). The fact that (weak) balance is broadly consistent
with the undirected representation of our network data, while status is more
consistent with the directed representation, shows that it possible for
different theories to be appropriate to different levels of resolution in the
representation of a single network.

In the final part of the paper, we describe further structural investigations
that provide insight into ways in which signed links are used in these
applications. First, we find that aspects of the theory of balance hold more
strongly on the subset of links in these networks that are {\em reciprocated}
--- consisting of directed links in both directions between two users. This
suggests that reciprocal link formation may follow a different pattern of use
in these systems than unreciprocated link formation. However, it is important
to note that such reciprocal relations account for only a small proportion of
the links between people on these sites.

Second, we find a connection between the sign of a link and the extent to which
it is {\em embedded} \cite{granovetter-embeddedness}, i.e., with the two
endpoints having links to many common neighbors.  A link is significantly more
likely to be positive when its two endpoints have multiple neighbors (of either
sign) in common. This observation is consistent with qualitative notions of
social capital \cite{burt-social-capital,coleman-social-capital} --- users with
common neighbors have relations that are ``on display'' in a social sense, and
hence have greater implicit pressure to remain positive.  Indeed in the three
different social applications that we study, this effect is strongest in the
case of voting for Wikipedia admins, which is the setting that makes the
relations most prominently visible to users. This suggests some of the ways in
which the presence of common neighbors, and more overt forms of public display,
can have an effect on the use of signed links.

These findings about aggregate structural properties also begin to address a
broad and largely open issue, which is to understand the sources of individual
variation in linking behavior.  While reciprocation and embeddedness are only
two dimensions along which to explore such variation, we believe that the
definitions and analysis pursued here can help in framing further investigation
of questions regarding individual variation.

\section{Related Work} \label{sec:related}
There is by now a large and rapidly growing literature on the analysis of
social networks arising in on-line domains \cite{newman-sirev}; as we noted at
the outset, this line of work has almost exclusively treated networks as
implicitly having positive signs only. For example, portions of our analysis
can be viewed as variants on the problem of {\em link prediction}
\cite{liben-nowell-link-pred} and {\em tie-strength prediction}
\cite{gilbert-tie-strength}, but in each case adapted to take the signs of
links into account.

Two recent papers in the analysis of on-line social networks stand out as
taking the signs of links into account. Brzozowski et al. study the positive
and negative relationships that exist on ideologically oriented sites such as
Essembly \cite{brzozowski-friend-foe}, but with the goal of predicting
outcomes of group votes rather than the broader organization of the social
network. Kunegis et al. study the friend/foe relationships on Slashdot, and
compute global network properties \cite{kunegis-slashdot-zoo}, but do not
evaluate theories of balance and status as we do here.

There are also large bodies of work involving negative relationships in on-line
domains that pursue directions different from our network focus here. One line
of work focuses on norms to control deviant behavior in on-line communities
(e.g. \cite{cosley-norms} and the references therein). In a different
direction, a large body of recent work in {\em sentiment analysis}
\cite{pang-lee-sentiment-book} has studied on-line textual data in which
individuals can express both positive and negative attitudes toward one
another, but without addressing the consequences for network structure.

The datasets we study here have also been investigated by researchers for other
purposes. Guha et al. study the trust network of Epinions \cite{guha-trust}.
Lampe et al. study the user rating mechanisms on Slashdot
\cite{lampe-slashdot}. Burke and Kraut study the voting process that produces
our Wikipedia signed network  \cite{burke-kraut-wikip-promote}, but with the
goal of modeling election outcomes.

Finally, the notion of status plays a role in many lines of work in the social
sciences, such as the role that behavior-status theory plays in social exchange
theory \cite{fisek-status,willer-exchange-book}. However, these notions are
distinct from the ways in which we formulate definitions of status as a
counterpart to balance in signed directed networks.

\section{Dataset description}

As described above, we consider three large online social networks where links
are explicitly positive or negative: (i) the trust network of the Epinions
product review Web site, where users can indicate their trust or distrust of
the reviews of others; (ii) the social network of the blog Slashdot, where a
signed link indicates that one user likes or dislikes the comments of another;
and (iii) the voting network of Wikipedia, where a signed link indicates a
positive or negative vote by one user on the promotion to admin status of
another.

\begin{table}[t]
\begin{center}
\small
\begin{tabular}{l||r|r|r}
   & Epinions & Slashdot & Wikipedia \\ \hline
  Nodes & 119,217 & 82,144 & 7,118 \\
  Edges & 841,200 & 549,202 & 103,747 \\
  $+$ edges & 85.0\% & 77.4\% & 78.7\% \\
  $-$ edges & 15.0\% & 22.6\% & 21.2\% \\
  Triads & 13,375,407 & 1,508,105 & 790,532
\end{tabular}
\end{center}
\vspace{-3mm} \caption{Dataset statistics.} \label{tab:datasets} \vspace{-3mm}
\end{table}

Table~\ref{tab:datasets} gives statistics for all three datasets. Our networks
have on the approximate order of tens to hundreds of thousand nodes, and less
than a million edges. In each network the edges are inherently directed, since
we know which user created the edge. In all networks the background proportion
of positive edges is about the same, with roughly 80\% of the edges having a
positive sign.

\begin{table}[t]
\begin{center}
\small
\begin{tabular}{l|l}
  Symbol & Meaning\\ \hline
  $T_i$ & Signed triad, also the number of triads of type $T_i$ \\
  $\Delta$ & Total number of triads in the network\\
  $p$ & Fraction of positive edges in the network \\
  $p(T_i)$ & Fraction of triads $T_i$, $p(T_i)=T_i/\Delta$ \\
  $p_0(T_i)$ & A priori prob. of $T_i$ (based on sign distribution)\\
  $E[T_i]$ & Expected number of triads $T_i$, $E[T_i]=p_0(T_i)\Delta$\\
  $s(T_i)$ & Surprise, $s(T_i) = (T_i-E[T_i])/\sqrt{\Delta p_0(T_i) (1-p_0(T_i))}$ \\
\end{tabular}
\end{center}
\vspace{-3mm} \caption{Table of symbols.} \label{tab:symbols} \vspace{-2mm}
\end{table}

\section{Analysis of Undirected Networks} \label{sec:undir}
We begin by analyzing the network data in an undirected representation, where
we do not take the directions of links into account. In this context, we can
evaluate the predictions of structural balance theory by considering the
frequencies of different types of signed {\em triads} --- sets of three nodes
with signed edges among all pairs.

Table~\ref{tab:undir} gives the counts of the four possible signed undirected
triads, while Table~\ref{tab:symbols} summarizes the symbols we use throughout
the paper. Let $p$ denote the fraction of positive edges in the network. The
four possible signed undirected triads are denoted $T_0, T_1, T_2$, and $T_3$
(Figure~\ref{fig:triads}). Among all triads in the data, the number that are of
type $T_i$ is denoted $|T_i|$ and the fraction of type $T_i$ is denoted
$p(T_i)$. Now, we would like to compare how this empirical frequency of triad
types compares to the corresponding frequencies if edge signs were produced at
random from the same background distribution of positive and negative signs.
Thus, we shuffle the signs of all edges in the graph (keeping the fraction $p$
of positive edges the same), and we let $p_0(T_i)$ denote the expected fraction
of triads that are of type $T_i$ after this shuffling.

If $p(T_i) > p_0(T_i)$, then triads of type $T_i$ are overrepresented in the
data relative to chance; if $p(T_i) < p_0(T_i)$, then they are
underrepresented. We also want to measure how significant this over- or
underrepresentation is. Thus, we define the \emph{surprise} $s(T_i)$ to be the
number of standard deviations by which the actual quantity of type-$T_i$
triads differs from the expected number under the random-shuffling
model.

\begin{table}[t]
  \begin{center}
  \small
  \begin{tabular}{l|l||r|r|r|r}
      \multicolumn{2}{c||}{Triad $T_i$ } & $|T_i|$ & $p(T_i)$ & $p_0(T_i)$ & $s(T_i)$\\
      \hline \hline
      \multicolumn{6}{c}{\bf Epinions} \\ 
      $T_3$ & $+$ $+$ $+$ & 11,640,257 & 0.870 & 0.621 & 1881.1 \\
      $T_1$ & $+$ $-$ $-$ & 947,855 & 0.071 & 0.055 & 249.4 \\
      $T_2$ & $+$ $+$ $-$ & 698,023 & 0.052 & 0.321 & -2104.8 \\
      $T_0$ & $-$ $-$ $-$ & 89,272 & 0.007 & 0.003 & 227.5 \\
      \multicolumn{6}{c}{\bf Slashdot} \\ \hline
      $T_3$ & $+$ $+$ $+$ & 1,266,646 & 0.840 & 0.464 & 926.5 \\
      $T_1$ & $+$ $-$ $-$ & 109,303 & 0.072 & 0.119 & -175.2 \\
      $T_2$ & $+$ $+$ $-$ & 115,884 & 0.077 & 0.406 & -823.5 \\
      $T_0$ & $-$ $-$ $-$ & 16,272 & 0.011 & 0.012 & -8.7 \\
      \multicolumn{6}{c}{\bf Wikipedia} \\ \hline
      $T_3$ & $+$ $+$ $+$ & 555,300 & 0.702 & 0.489 & 379.6 \\
      $T_1$ & $+$ $-$ $-$ & 163,328 & 0.207 & 0.106 & 289.1 \\
      $T_2$ & $+$ $+$ $-$ & 63,425 & 0.080 & 0.395 & -572.6 \\
      $T_0$ & $-$ $-$ $-$ & 8,479 & 0.011 & 0.010 & 10.8 \\
  \end{tabular}
  \end{center}
  \caption{Number of balanced and unbalanced undirected triads.
  }
  \label{tab:undir}
  \vspace{-5mm}
\end{table}

Due to the Central Limit Theorem the distribution of $s(T_i)$ is approximately
a standard normal distribution and so we would expect surprise on the order of
tens to already be significant ($s(T_i)=6$ gives 
a p-value of $\approx 10^{-8}$).
However, the values of surprise we find in our data are typically much larger.
This means that due to the scale of the data and the large number of triads
almost all our observations are statistically significant with p-values
practically equal to zero.

We find that the all-positive triad $T_3$ is heavily overrepresented in all
three datasets, and the triad $T_2$ consisting of two enemies with a common
friend is heavily underrepresented. Based on the relative magnitudes of
$p(T_i)$ and $p_0(T_i)$, we see that $T_3$ tends to be over represented by
about 40\% in all three datasets. Similarly, the unbalanced triad $T_2$ is
underrepresented by about 75\% in Epinions and Slashdot and 50\% in Wikipedia.
These observations so far fit well into Heider's original notion of structural
balance.

However, the relative abundances of triad types $T_1$ (single positive edge)
and $T_0$ (all negative edges) differ between the datasets, and none of the
datasets follow Heider's theory in both having $T_1$ overrepresented and $T_0$
underrepresented. Thus, the picture is more consistent with Davis's weaker
notion of balance, where $T_2$ is viewed as implausible but there is no {\em a
priori} reason to favor one of $T_1$ or $T_0$ over the other.

\section{Analysis of Evolving Directed Networks}
\label{sec:evol}
\def\T{{t}}
\def\DS{s_{g}}
\def\SS{s_{r}}
\def\Bo{B_{g}}
\def\Bi{B_{r}}
\def\So{S_{g}}
\def\Si{S_{r}}

We now consider the networks in these systems as directed graphs, incorporating
the fact that the links being created go from one user to another, with the
sign of a link from $A$ to $B$ being generated by $A$. In the introduction, we
discussed how the theories of balance and status offer competing
interpretations for how we should expect such directed links to be signed. For
example, as noted there, {\em positive cycles} --- that is, directed triads
with positive links from $A$ to $B$ to $C$ to $A$ --- are underrepresented in
the data. This conflicts with balance theory, but is consistent with status
theory.

\xhdr{Timing and Diversity: Generative and Receptive Baselines.} Beyond just
the directionality of links, there are additional features of the data that we
take into account when evaluating these models. First, links are created at
specific points in time, so rather than thinking of directed triads as existing
in a static snapshot of the network, we consider the order in which links are
added to the network. Thus, we study how directed triads form, as follows. When
a user $A$ links to a user $B$, suppose there is already a user $X$ with the
property that $X$ has links to or from $A$, and also to or from $B$.  This
means there is a two-step {\em semi-path} from $A$ to $B$ through $X$ (a path
in which the directions of the edges do not matter), and the formation of the
$A$-$B$ link adds a short-cut to this path, producing a directed triad on $A$,
$B$, and $X$.

Second, different users make use of positive and negative signs differently. At
the most basic level, some users produce links almost exclusively of one sign
or the other, while others produce a relatively even mix of both positive and
negative links. We will refer to the overall fraction of positive signs that a
user creates, considering all her links, as her {\em generative baseline}.
Similarly, some users receive links that are almost exclusively of one sign or
the other, while others receive a mix of signs. We will refer to the overall
fraction of positive signs in the links a user receives as his {\em receptive
baseline}. Given this, we should compare the abundance of positive and negative
links to the generative and receptive baselines of the users producing and
receiving these links.

Once we incorporate these aspects of the data, we discover further mysteries
--- beyond just the scarcity of positive cycles --- that seem to call for
alternatives to balance theory. For example, consider the case of {\em joint
positive endorsement} --- a situation in which a node $X$ links positively to
each of two nodes $A$ and $B$. Suppose that in this case, $A$ now forms a link
to $B$ (i.e., triad $\T_9$ of Figure~\ref{fig:evolEpin}); should we expect
there to be an elevated probability of the link being positive, or a reduced
probability of the link being positive?

In fact, in our data, the question turns out to have a more subtle answer than
either of these alternatives. The link that is produced in this situation is
{\em more likely} to be positive than the generative baseline of $A$, but at
the same time {\em less likely} to be positive than the receptive baseline of
$B$. Balance theory, of course, makes a much more naive prediction: since $A$
and $B$ are both friends of $X$, they should be friends of each other. Can
status theory explain this dual and opposite pair of deviations from the
baselines of $A$ and $B$?

We now show that in fact it can, and explaining how this works forms the
motivation for a theory of how status effects can influence the signs of
directed links.

\subsection{Formulating a Theory of Status}

Since the phenomenon we are trying to capture is subtle but in the end familiar
from everyday life, we begin with a hypothetical example to motivate the
subsequent definitions.

\xhdr{A Motivating Example.} Suppose we were to interview the players on a
college soccer team: for certain players $A$, and certain teammates $B$ of $A$,
we ask, ``How do you think the skill of player $B$ compares to yours?'' Suppose
further that the players roughly agree on a ranking of each other by skill,
which serves as an approximate (though not perfect) ranking of the team members
by status. From the results of these interviews, we could produce a signed
directed graph whose nodes are the players, and with a directed edge from $A$
to $B$ if we asked $A$ for her opinion of $B$.  A positive link from $A$ to $B$
would indicate that $A$ thinks highly of $B$'s skill relative to her own, while
a negative link would indicate that $A$ thinks she is better than $B$.

If we were just given this signed directed graph, and knew nothing else about
the soccer team, then we could still make inferences about the signs of links
that we haven't yet observed, using the {\em context} provided by the rest of
the network. Suppose for example that we are about to ask player $A$'s opinion
of another player $B$, but we don't currently have $A$'s answer and hence don't
yet know the sign of the link from $A$ to $B$. We can nonetheless make
predictions about it from the links whose signs we do know, as follows. Suppose
that we know from the data already collected that $A$ and $B$ have each
received a positive evaluation from a third player $X$. Here is a pair of facts
we could conjecture about the link from $A$ to $B$, given the positive links
from $X$ to $A$ and $B$.
\begin{itemize}
  \denselist
  \item Since $B$ has been positively evaluated by another team member,
      $B$ is more likely than not to have above-average skill. Therefore,
      the evaluation that $A$ gives $B$ should be more likely to be
      positive than an evaluation given by $A$ to a random team member.
  \item Since $A$ has been positively evaluated by another team member,
      $A$ is also more likely than not to have above-average skill.
      Therefore, the evaluation that $A$ gives $B$ should be less likely
      to be positive than an evaluation received by $B$ from a random team
      member.
\end{itemize}

There are several subtleties here. First, we're using the indirection provided
by a third party $X$ to make inferences about the relation between $A$ and $B$,
based on assumptions about status. Second, the context provided by $X$ causes
the sign of the $A$-$B$ link to deviate from a random baseline in {\em
different} directions depending on whether we're looking at it from $A$'s point
of view or $B$'s point of view. More precisely, since $B$ has above-average
skill, $A$ will likely give $B$ a higher evaluation than $A$ would give to a
random team member. On the other hand, since $A$ has above-average skill, $B$
is less likely to receive a positive evaluation from $A$ than she would receive
from a random team member. Despite the complexity of these conclusions, they
reflect genuine and natural properties of status ordering among a group of
people. They also agree with our observations about joint positive endorsement
in the data mentioned above.

We turn now to the data, where we will find that the users of these on-line
networks create signed links in ways that correspond closely to the behavior
of the players on our hypothetical soccer team. But extracting this finding
from the data will require formulating a sequence of definitions that captures
the intuition suggested by this example.

\begin{figure}[!t]
  \begin{center}
  \begin{tabular}{cccc}
  $\T_1$ & $\T_2$ & $\T_3$ & $\T_4$ \\
  \includegraphics[scale=0.46]{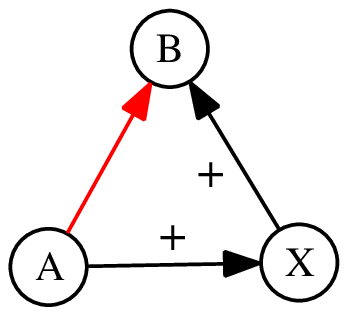} &
  \includegraphics[scale=0.46]{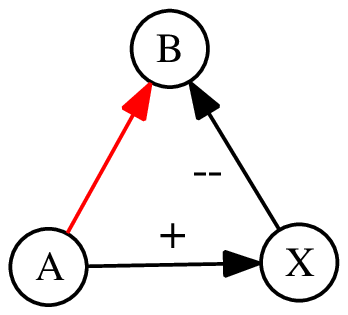} &
  \includegraphics[scale=0.46]{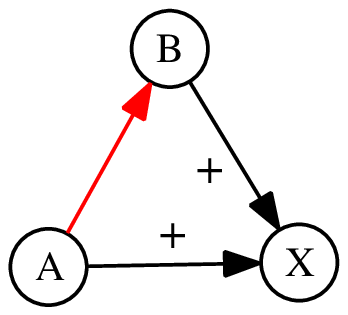} &
  \includegraphics[scale=0.46]{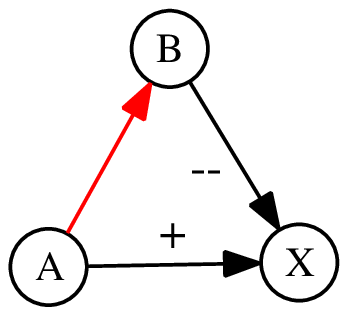} \\
  $\T_5$ & $\T_6$ & $\T_7$ & $\T_8$ \\
  \includegraphics[scale=0.46]{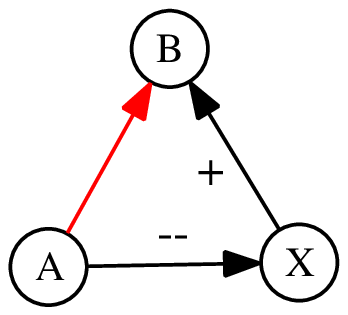} &
  \includegraphics[scale=0.46]{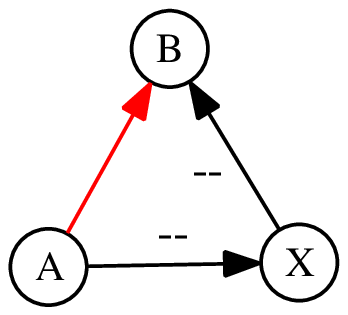} &
  \includegraphics[scale=0.46]{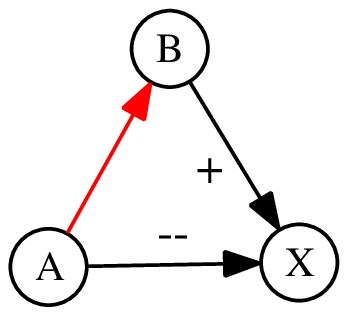} &
  \includegraphics[scale=0.46]{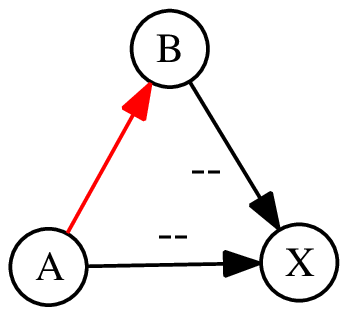} \\
  $\T_9$ & $\T_{10}$ & $\T_{11}$ & $\T_{12}$ \\
  \includegraphics[scale=0.46]{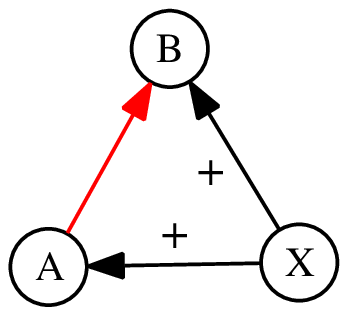} &
  \includegraphics[scale=0.46]{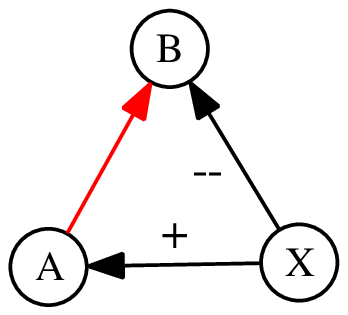} &
  \includegraphics[scale=0.46]{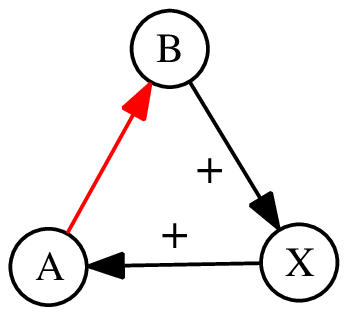} &
  \includegraphics[scale=0.46]{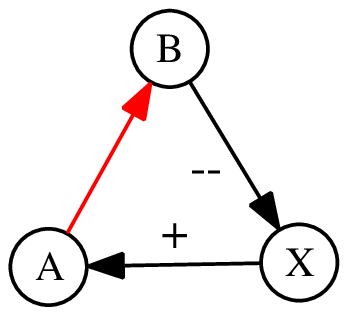} \\
  $\T_{13}$ & $\T_{14}$ & $\T_{15}$ & $\T_{16}$ \\
  \includegraphics[scale=0.46]{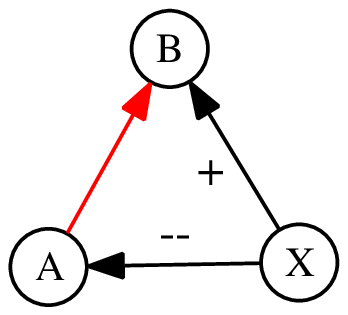} &
  \includegraphics[scale=0.46]{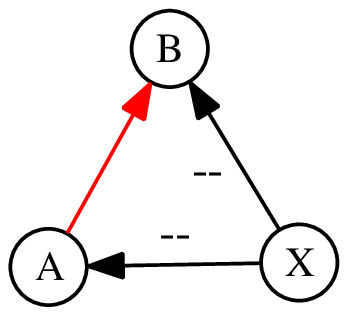} &
  \includegraphics[scale=0.46]{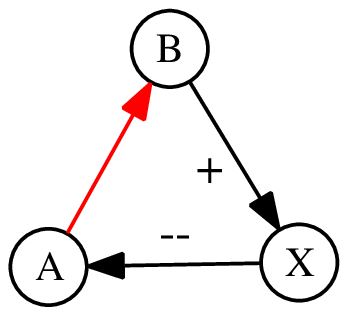} &
  \includegraphics[scale=0.46]{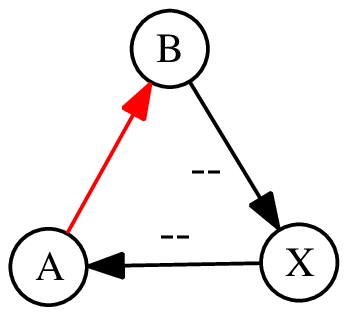} \\
  \end{tabular}
  \small
    \begin{tabular}{l|rrrr|rrrr}
  $\T_i$ & count & $P(+)$ & $\DS$ & $\SS$ & $\Bo$ & $\Bi$ & $\So$ & $\Si$ \\ \hline
  $\T_1$ & 178,051 & 0.97 & 95.9 & 197.8 & $\checkmark$ & $\checkmark$ & $\checkmark$ & $\checkmark$ \\
  $\T_2$ & 45,797 & 0.54 & -151.3 & -229.9 & $\checkmark$ & $\checkmark$ & $\checkmark$ & $\circ$ \\
  $\T_3$ & 246,371 & 0.94 & 89.9 & 195.9 & $\checkmark$ & $\checkmark$ & $\circ$ & $\checkmark$ \\
  $\T_4$ & 25,384 & 0.89 & 1.8 & 44.9 & $\circ$ & $\circ$ & $\checkmark$ & $\checkmark$ \\
  $\T_5$ & 45,925 & 0.30 & 18.1 & -333.7 & $\circ$ & $\checkmark$ & $\checkmark$ & $\checkmark$ \\
  $\T_6$ & 11,215 & 0.23 & -15.5 & -193.6 & $\circ$ & $\circ$ & $\checkmark$ & $\checkmark$ \\
  $\T_7$ & 36,184 & 0.14 & -53.1 & -357.3 & $\checkmark$ & $\checkmark$ & $\checkmark$ & $\checkmark$ \\
  $\T_8$ & 61,519 & 0.63 & 124.1 & -225.6 & $\checkmark$ & $\circ$ & $\checkmark$ & $\checkmark$ \\
  $\T_9$ & 338,238 & 0.82 & 207.0 & -239.5 & $\checkmark$ & $\circ$ & $\checkmark$ & $\checkmark$ \\
  $\T_{10}$ & 27,089 & 0.20 & -110.7 & -449.6 & $\checkmark$ & $\checkmark$ & $\checkmark$ & $\checkmark$ \\
  $\T_{11}$ & 35,093 & 0.53 & -7.4 & -260.1 & $\circ$ & $\circ$ & $\checkmark$ & $\checkmark$ \\
  $\T_{12}$ & 20,933 & 0.71 & 17.2 & -113.4 & $\circ$ & $\checkmark$ & $\checkmark$ & $\checkmark$ \\
  $\T_{13}$ & 14,305 & 0.79 & 23.5 & 24.0 & $\circ$ & $\circ$ & $\checkmark$ & $\checkmark$ \\
  $\T_{14}$ & 30,235 & 0.69 & -12.8 & -53.6 & $\circ$ & $\circ$ & $\checkmark$ & $\circ$ \\
  $\T_{15}$ & 17,189 & 0.76 & 6.4 & 24.0 & $\circ$ & $\circ$ & $\circ$ & $\checkmark$ \\
  $\T_{16}$ & 4,133 & 0.77 & 11.9 & -2.6 & $\checkmark$ & $\circ$ & $\checkmark$ & $\circ$ \\ \hline \hline
  \multicolumn{5}{r|}{Number of correct predictions} & 8 & 7 & 14 & 13 \\
       \end{tabular}
       \vspace{-3mm}
  \caption{Top: All contexts $(A,B;X)$.
  Red edge is the edge that closes the triad.
  Bottom:
  Surprise values and predictions based on the competing theories of structural balance
  and status. $\T_i$ refers to triad contexts above; {\em Count}: number of contexts $\T_i$;
  $P(+)$: prob. that closing red edge is positive; $\DS$: surprise of edge initiator giving a positive edge;
  $\SS$: surprise of edge destination receiving a positive edge;
$\Bo$: consistency of balance with generative surprise;
$\Bi$: consistency of balance with receptive surprise;
$\So$: consistency of status with generative surprise;
$\Si$: consistency of status with receptive surprise.
}
       \vspace{-5mm}
         \label{fig:evolEpin}
  \end{center}
\end{figure}

\xhdr{Contextualized Links.} The first portion of our definitions capture the
idea that we will evaluate the sign of a link created from $A$ to $B$ in the
{\em context} of $A$ and $B$'s relations to additional nodes $X$ with whom they
have links. (For example, the node $X$ in our example who jointly endorses $A$
and $B$.) Thus, we define a {\em contextualized link} (more briefly, a {\em
c-link}) to be a triple $(A,B;X)$ with the the property that a link forms from
$A$ to $B$ {\em after} each of $A$ and $B$ already has a link either to or from
$X$. Overall there are sixteen different types of c-links, as the edge between
$X$ and $A$ can go in either direction and have either sign yielding four
possibilities, and similarly for the edge between $X$ and $B$, for a total of 
$4 \cdot 4=16$. For each of these types of c-links we are interested in the
frequencies of positive versus negative labels for the edge from $A$ to $B$.
Figure~\ref{fig:evolEpin} shows all the possible types of c-links, labeled
$\T_1$--$\T_{16}$.

Now, for a particular type of c-link, we look at the set of all c-links
$(A,B;X)$ of this type, and ask: what fraction of the links from $A$ to $B$ in
this set are positive? Moreover, how does this fraction compare to what one
would expect from the generative baselines of the nodes $A$ and the receptive
baselines of the nodes $B$ that are involved in the creation of these $A$-$B$
links? If we can quantify the answer to this question in our data, we can look
for effects like we saw in our motivating example --- there, in the case of
positive links from $X$ to $A$ and $B$, we believed the likelihood of a positive
$A$-$B$ edge should exceed the generative baseline of $A$ but should lie below
the receptive baseline of $B$.

Let's consider a particular type $\T$ of c-link, and suppose that $(A_1, B_1;
X_1), (A_2, B_2; X_2), \ldots, (A_k, B_k; X_k)$ is a list of all instances of
this type $\T$ of c-link in our data. We define the {\em generative baseline}
for this type $\T$ to be the sum of the generative baselines of all nodes
$A_i$. This quantity is simply the expected number of positive edges we would
get {\em if we let each $A_i$-$B_i$ link form according to the generative
baseline of $A_i$}. We then define the {\em generative surprise} $\DS(\T)$ for
this type $\T$ to be the (signed) number of standard deviations by which the
actual number of positive $A_i$-$B_i$ edges in the data differs above or below
this expectation. In other words, if the context provided by the node $X$ and
its links with $A$ and $B$ had no effect on the sign of the $A$-$B$ link being
formed, so that each node $A_i$ simply drew the sign of her link to $B_i$
according to her generative baseline, then we should expect to see 
a generative surprise of $0$ for this type $\T$.

We set up the corresponding definitions for the nodes $B_i$ as the recipients
of the links. We define the {\em receptive baseline} for this type $\T$ of
c-link to be the sum of the receptive baselines of all nodes $B_i$, and we
define the {\em receptive surprise} $\SS(\T)$ to be the (signed) number of
standard deviations by which the actual number of positive $A_i$-$B_i$ edges in
the data differs above or below this expectation.

\xhdr{Incorporating the Role of Status.} Finally, we bring the role of status
into this theory. For this, it is useful to return once more to our motivating
example. When a player $X$ on our hypothetical soccer team gave positive
evaluations to both $A$ and $B$, we concluded --- in the absence of any further
information --- that $A$ and $B$ were likely to have above-average status. We
would have concluded the same thing had $A$ and $B$ given negative evaluations
to $X$. On the other hand, if $X$ had evaluated $A$ and $B$ negatively, or had
they evaluated $X$ positively, then we should have concluded that $A$ and $B$
were more likely than not to have below-average status.

This reasoning provides a way to assign status values to $A$ and $B$ in any
type of c-link, as follows. We first assign the node $X$ a status of $0$.
Then, if $X$ links positively to $A$, or $A$ links negatively to $X$, we assign
$A$ a status of $1$; otherwise, we assign $A$ a status of $-1$. We use the same
rule for assigning a status of $1$ or $-1$ to $B$. Thus we say that the
generative surprise for type $t$ is {\em consistent with status} if $B$'s
status has the same sign as the generative surprise: in this case, high-status
recipients $B$ receive more positive evaluations than would be expected from
the generative baseline of the node $A$ producing the link. We say that the
receptive surprise for type $t$ is consistent with status if $A$'s status has
the opposite sign from the receptive surprise: high-status generators of
links $A$ produce fewer positive evaluations than would be expected from the
receptive baseline of the node $B$ receiving the link.

\subsection{Results}

We now evaluate the predictions of these theories on the two networks, Epinions
and Wikipedia, for which we have data on the exact order in which the links
were created. We focus our discussion on Epinions, for which the data is an
order of magnitude larger; the results are quite similar on the smaller
Wikipedia dataset, with differences that we note below.

We consider four theories to explain the signs of the links that are produced.
The first two are the consistency of status with generative  and
receptive surprise, as just defined. The other two theories are the analogous
forms of consistency with Heider's original notion of balance. Specifically, we
say that Heider balance is consistent with generative surprise for a particular
c-link type if the sign of the generative surprise is equal to the sign of the
edge as predicted by balance. Analogously, we say that Heider balance is
consistent with receptive surprise for a particular c-link type if the sign of
the receptive surprise is equal to the sign of the edge as predicted by
balance.

We find that the predictions of status with respect to both generative  and
receptive surprise perform much better against the data that the predictions of
structural balance. Indeed, status is consistent with generative  and
receptive surprise on the vast majority of c-link types; as shown in
Figure~\ref{fig:evolEpin}, it is consistent on 14 and 13 types respectively.
This includes the case of joint endorsement (type $\T_9$ in
Figure~\ref{fig:evolEpin}) --- which is in fact the most abundant type of
c-link in the data --- and also includes the natural counterpart of joint
endorsement, in which $A$ and $B$ each link negatively to $X$ (type $\T_8$). It
also includes the case of a positive cycle (type $\T_{11}$), discussed earlier
as well.\footnote{On the Wikipedia dataset, the results for receptive surprise
are almost identical; status is consistent with receptive surprise on all
c-link types except for the same three exceptional cases as Epinions, $\T_2$,
$\T_{14}$, and $\T_{16}$, and one more: $\T_4$. We find this close alignment
quite surprising given the very different kinds of activities that the Epinions
and Wikipedia links represent. On Wikipedia, status is also consistent with
generative surprise on 12 of the 16 triad types, though here the types where
there is inconsistency differ more from Epinions: $\T_{14}$ (as in Epinions),
$\T_5$, $\T_8$, and $\T_{16}$.}

Structural balance is a much weaker fit to the data: balance is consistent with
generative surprise for only 8 of the 16 types of c-links, and consistent with
receptive surprise for only 7 of the 16. We also evaluated consistency of
generative  and receptive surprise with respect to Davis's weaker notion of
balance, with similar results. The one subtlety in evaluating the data with
respect to Davis balance is that Davis's theory does not predict the sign of
the $A$-$B$ edge in c-link types where the two existing edges with $X$ are both
negative ($\T_6, \T_8, \T_{14}$, and $\T_{16}$): for these triads, either
a positive or a negative $A$-$B$ link would be consistent with Davis's theory,
and so no prediction can be made. Thus, we evaluate consistency of Davis
balance with respect to generative and receptive surprise only on the
remaining 12 c-link types; here, we find consistency in 6 and 7 of the 12 cases
respectively. This too is much weaker than the predictions of status.

We also consider the structure of the cases in which status theory fails to
make a correct prediction, analyzing the possible strengthenings of the theory
that this might hint at. First, we observe that one of the two c-link types
where status is inconsistent with generative surprise is the configuration in
which $A$ and $B$ each link positively to $X$ (type $t_3$). This is one of the
most basic settings for structural balance in Heider's work: if two people each
like a third party, then one should expect them to have positive relations. It
thus suggests where users of these systems may be relying on balance-based
reasoning more than status-based reasoning.

We can get further insights from the cases where status theory is inconsistent
with the data. In particular, the 16 c-link types can be divided into four
groups of four each, based on whether $A$ has high or low status relative to
$X$, and whether $B$ has high or low status relative to $X$. In looking at
where status theory makes mistakes, it is almost exclusively on the c-link
types where $A$ and $B$ are both posited to have {\em low} status relative to
$X$. This corresponds to the types $\T_2$, $\T_3$, $\T_{14}$, and $\T_{15}$; we
observe that with respect to generative surprise, both of status theory's
mistakes occur on types of this form, and with respect to receptive surprise,
two of status theory's three mistakes occur on types of this form.

Even further, the mistakes of status with respect to generative  and
receptive surprise on these types constitute natural ``duals'' to each other.
Note first that if we reverse both the direction and the sign of an edge, we
preserve the status relation of the two endpoints (e.g. a positive link from
$A$ to $X$ or a negative link from $X$ to $A$ both suggest that $A$ has lower
status than $X$). With this in mind, we observe that if we take the types
$\T_3$ and $\T_{15}$ on which status theory makes its two mistakes with respect
to generative surprise, and we reverse the directions and signs of both edges
involving $X$, we get the c-link types $\T_2$ and $\T_{14}$ --- these are the
other two c-link types where $A$ and $B$ have low status relative to $X$, and
they are two of the three types on which status theory makes mistakes with
respect to receptive surprise.

It is thus natural to conjecture that the use of signed links deviates most
strongly from status theory when $A$ is predicted to impute low status to both
herself and $B$. Now that this behavioral asymmetry has been identified in the
data, via our formulation of this theory, developing a more refined theory of
status that takes this asymmetry into account is an interesting direction for
further work.

\section{Reciprocation of Directed Edges}

\begin{table}[t]
  \begin{center}
  \small
  \begin{tabular}{l|r|r}
    {\bf Epinions} & Count & Probability \\ \hline \hline
    $P(+|+)$  &  38,415 & 0.969\\
    $P(-|+)$  &  1,204  & 0.031 \\
    $P(+|-)$  &  1,192  & 0.692 \\
    $P(-|-)$  &  560    & 0.308 \\ \hline
    {\bf Wikipedia} & Count & Fraction \\ \hline \hline
    $P(+|+)$  &  2,509 & 0.945 \\
    $P(-|+)$  &  145   & 0.055 \\
    $P(+|-)$  &  193   & 0.706 \\
    $P(-|-)$  &  80    & 0.294 \\
  \end{tabular}
  \end{center}
  \vspace{-3mm}
  \caption{Edge reciprocation. Given that the first edge was of sign $X$
$P(Y|X)$ give the probability that reciprocated edge is $Y$.}
  \label{tab:recip}
  \vspace{-5mm}
\end{table}

Thus far we have found that balance theory is a reasonable approximation to the
structure of signed networks when they are viewed as undirected graphs, while
status theory better captures many of the properties when the networks are
viewed in more detail as directed graphs that grow over time.

To understand the boundary between these two theories and where they apply, it
is interesting to consider a particular subset of these networks where the
directed edges are used to create symmetric relationships. This subset is the
collection of edges that are {\em reciprocal}: cases in which there are two
nodes $A$ and $B$ such that $A$ links to $B$ and $B$ also links to $A$. (If
the $B$-$A$ link forms after the $A$-$B$ link, we say that $B$ {\em
reciprocates} the link to $A$.) In our data, only about 3-5\% of the edges
represent the reciprocation of an existing link, so this is far from being a
dominant mode of link creation on these systems. But it is an interesting
mode of link creation, in that it represents a directly mutual relationship
between two individuals $A$ and $B$, which is the setting in which balance
theory has been more relevant to our earlier analyses.

Our findings for this type of linking
suggest the following intuitively natural picture: in the
relatively small portion of these networks where mutual back-and-forth
interaction takes place, the principles of balance are more pronounced than
they are in the larger portions of the networks where signed linking (and hence
evaluation of others) takes place asymmetrically. In other words, users treat
each other differently in the context of back-and-forth interaction than when
they are using links to refer to others who do not link back.

We summarize the results in Table~\ref{tab:recip}.
First, we find that the
reciprocation of positive $A$-$B$ edges is closely consistent with balance
rather than status, while the reciprocation of negative edges seems to follow
a hybrid of the two principles. Specifically, if $A$ links positively to $B$,
then balance predicts that $B$ should link positively to $A$, while status
predicts that $B$ has the higher status and should therefore link negatively
to $A$. For the two systems in which we have data on the order of edge
creation --- Epinions and Wikipedia --- we find that the data clearly supports
the balance interpretation, as shown in Table~\ref{tab:recip}. When a $B$-$A$
link reciprocates a positive $A$-$B$ link, this $B$-$A$ link is positive well
over 90\% of the time --- much higher than the roughly 80\% fraction of
positive links in the system as a whole.

Reciprocation of a negative $A$-$B$ link, on the other hand, displays
ingredients of both theories. When $A$ links negatively to $B$ and $B$
subsequently links to $A$, balance theory predicts a negative link while
status theory predicts a positive one (since $A$ should have higher status).
In the data, such $B$-$A$ links are positive roughly 70\% of the time. This
shows that users respond to a negative link with a positive link a majority of
the time, but still at a rate below the
80\% fraction of positive links in the system as a whole, suggesting a
deviation in the direction of the balanced-based interpretation.

From Table~\ref{tab:recip}, it is also interesting to observe how similar the
probabilities for all kinds of reciprocation are between the two systems
Epinions and Wikipedia. This is particularly striking given how different the
level of public display of link signs is on these systems; it
suggests that these rates of alignment in the signs are being driven
by forces that may be relatively robust to the way in which link signs are
presented.

\subsection{The Role of Triadic Structure in Reciprocation}

\begin{table}[t]
  \small
  \begin{center}
  \begin{tabular}{l||r|r|r|r}
  {\bf Epinions} & Triads & $P(\textrm{RSS})$ & $P(+ | +)$ & $P(- | -)$ \\ \hline
  Balanced   & 348,538 & 0.929 & 0.941 & 0.688 \\
  Unbalanced & 74,860  & 0.788 & 0.834 & 0.676 \\ \hline \hline
  {\bf Wikipedia} & Triads & $P(\textrm{RSS})$ & $P(+ | +)$ & $P(- | -)$ \\   \hline 
  Balanced   & 53,973 & 0.912 & 0.934 & 0.336 \\
  Unbalanced & 13,542 & 0.661 & 0.878 & 0.195
  \end{tabular}
  \end{center}
  \vspace{-3mm}
  \caption{Edge reciprocation in balanced and unbalanced triads. {\em Triads:}
  number of balanced/unbalanced triads in the network where one of the edges was
  reciprocated. $P(\textrm{RSS})$: probability that the reciprocated edge is of
  the same sign. $P(+|+)$: probability that the $+$ edge is later reciprocated with a
  plus.
  $P(-|-)$: probability that the
  $-$ edge is reciprocated with a minus.}
  \vspace{-5mm}
  \label{tab:recip3to4}
\end{table}

We now consider how reciprocation between $A$ and $B$ is affected by the
context of $A$ and $B$'s relationships to third nodes $X$. Specifically,
suppose that an $A$-$B$ link is part of 
a directed triad in which each of $A$ and
$B$ has a link to or from a node $X$. Now, $B$ reciprocates the link to $A$. As
indicated in Table~\ref{tab:recip3to4}, we find that the $B$-$A$ link is
significantly more likely to have the same sign as the $A$-$B$ link when the
original triad on $A$-$B$-$X$ (viewed as an undirected triad) is structurally
balanced. In other words, when the initial $A$-$B$-$X$ triad is unbalanced,
there is more of a latent tendency for $B$ to ``reverse the sign'' when she
links back to $A$. The effect holds in all cases; it is more pronounced in
Wikipedia than in Epinions, which is interesting given the difference in how
public the edge signs are.

This result further indicates how balance-based effects seem to be at work
in the portions of the networks where directed edges point in both
directions, reinforcing mutual relationships. We conjecture that this tension
between mutuality and asymmetry in different parts of the network will be
relevant in understanding more deeply the interplay between status and balance
effects in shaping the formation of links.

\section{Further Structural Analysis of Signed Links}
\label{sec:struc}
Finally, we explore some additional connections between network structure and
the signs of links, focusing on the embeddedness of edges and on the subgraphs
consisting only of positive links and only of negative links.
For these structural results, we analyze the networks as undirected graphs.

\subsection{Embeddedness of positive and negative ties}

We begin by trying to characterize the parts of the network in which positive
ties are more likely to occur. Roughly, we find that positive ties are more
likely to be clumped together, while negative ties tend to act more like
bridges between islands of positive ties.

\begin{figure*}[t]
  \begin{center}
  \small
  \begin{tabular}{ccc}
    \hspace{3mm}\includegraphics[width=0.25\textwidth]{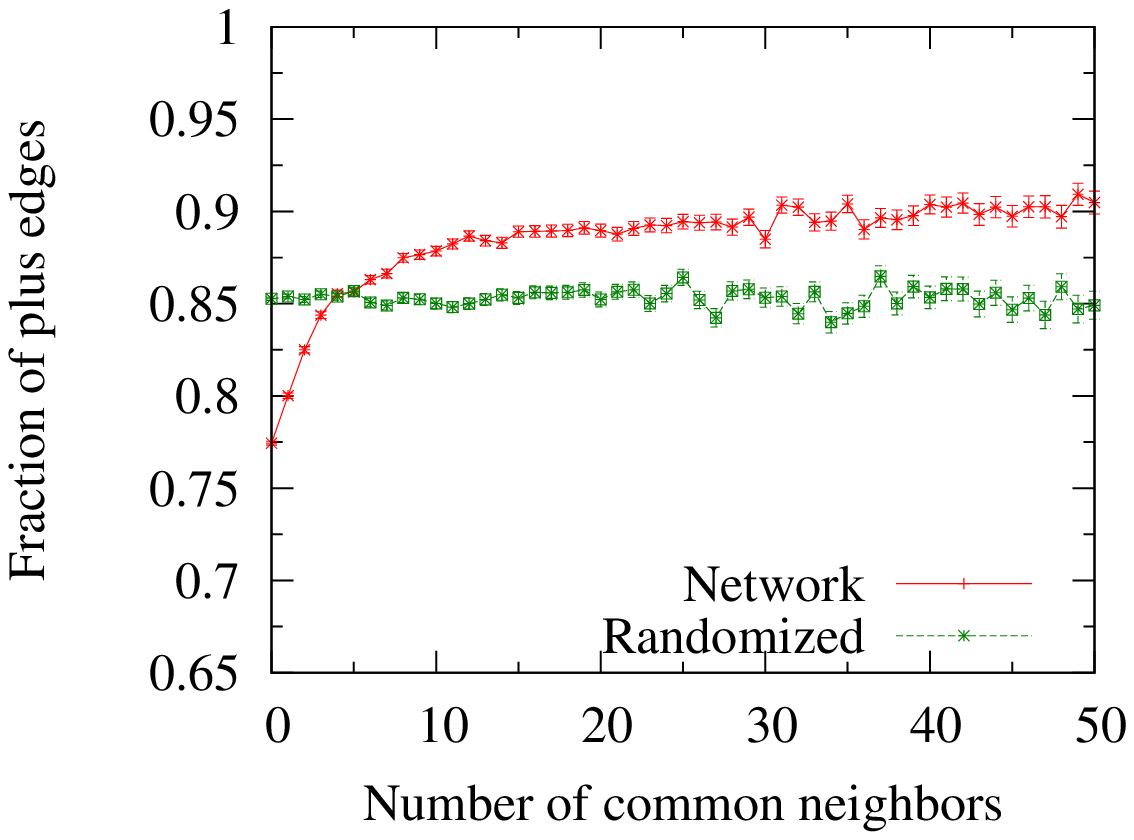} \hspace{3mm} &
    \hspace{3mm}\includegraphics[width=0.25\textwidth]{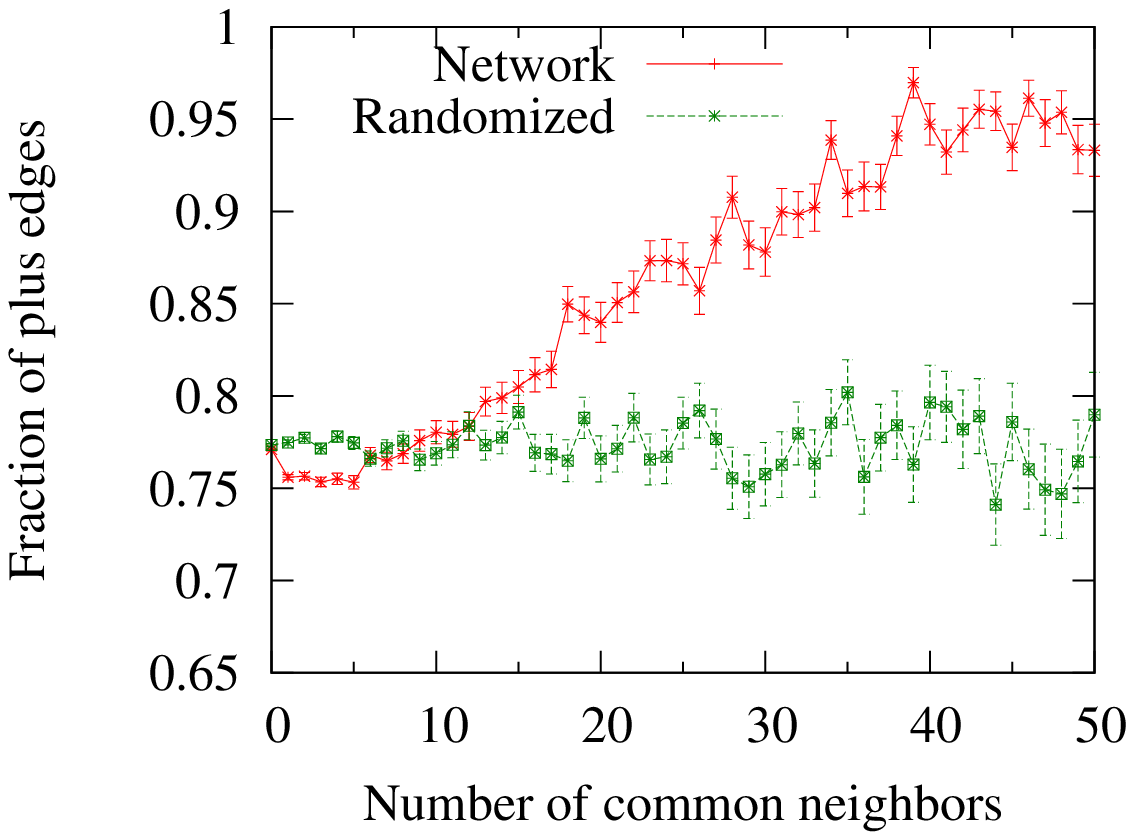} \hspace{3mm} &
    \hspace{3mm}\includegraphics[width=0.25\textwidth]{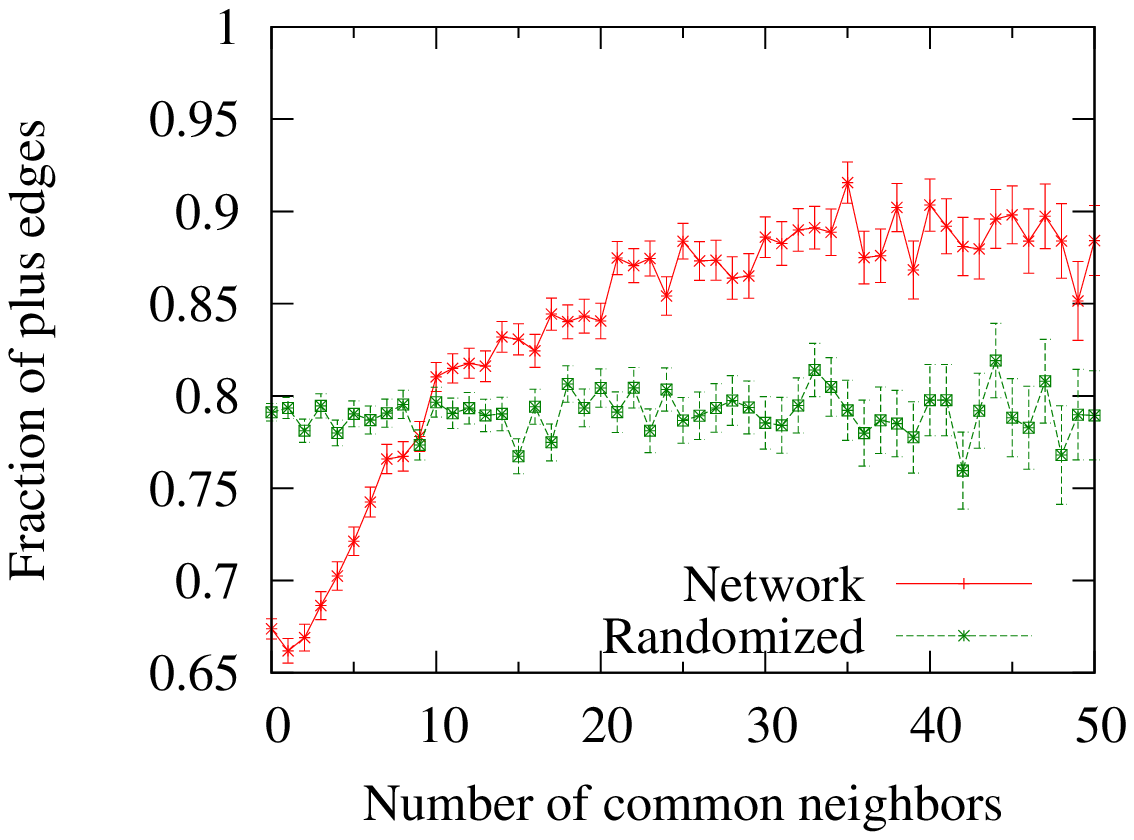} \hspace{3mm}\\
    (a) Epinions & (b) Slashdot & (c) Wikipedia \\
  \end{tabular}
    \vspace{-3mm}
    \caption{Embeddedness of positive ties in the network. More
    embedded edges tend to be more positive.}
    \label{fig:cmnnbrs}
    \vspace{-5mm}
  \end{center}
\end{figure*}

We explore this issue in Figure \ref{fig:cmnnbrs} by plotting the probability
that an edge is positive as a function of its {\em embeddedness} , i.e., the
number of common neighbors that its endpoints have
\cite{granovetter-embeddedness}, or equivalently, the number of distinct triads
the edge participates in. For each dataset we plot two curves. In green, we
show the results of a random-shuffling baseline --- the sign probability we
would get as a function of embeddedness if edge signs were determined randomly
and independently with probability $p$ for each edge. As is clear, there is no
dependence here between an edge's sign and its embeddedness, so the green curve
is approximately flat.

However, in the real data (red) we see a completely different picture. Edges
that are not well embedded (with endpoints having fewer than around 10 shared
neighbors) tend to be more negative than expected based on the background
probability $p$ of positive ties. However, as an edge is more embedded
(participating in more triads) it tends to be increasingly positive. That is, a
link is significantly more likely to be positive when its two endpoints have
multiple neighbors (of either sign) in common. These findings are consistent
across all three datasets. This suggests that positive edges tend to occur in
better embedded (densely linked) groups of nodes, while negative edges tend to
participate in fewer triangles, which indicates that they act as connections
between the well-embedded sets of positive ties.

As mentioned in the Introduction, this observation is not part of the
formulation of balance theory (and does not follow from it), but it is
consistent with the notion from social-capital theory of embedded edges being
more ``on display'' \cite{burt-social-capital,coleman-social-capital}.
Moreover, among our three datasets, this phenomenon is most pronounced for the
Wikipedia voting data.  This is also the only one of the three sites where the
social relations are explicitly displayed to a broad set of users --- thus
putting the relations even more highly on display. Thus these results are
particularly well explained in terms of implicit pressure to remain positive.

\subsection{All-Positive and All-Negative Networks}

To explore further the different roles played by positive and negative links in
these networks, we study the sub-networks composed exclusively of the positive
links and exclusively of the negative links. That is, we define the
all-positive network to be the subgraph consisting only of the positive links,
and the all-negative network to be the subgraph consisting only of the negative
links. We also compare these to randomized baselines, in which we first
randomly shuffle the edge signs in the full network, and then extract the
all-positive and all-negative networks from these shuffled versions.

\begin{table}[t]
  \small
  \begin{tabular}{l||r|r||r|r||r|r}
      &  \multicolumn{2}{c||}{Size} &  \multicolumn{2}{c||}{Clustering}  &  \multicolumn{2}{c}{Component}  \\
      &  Nodes & Edges &  Real & Rnd  &  Real & Rnd  \\ \hline
      \hline
    Epinions: $-$  &  119,090 & 123,602 & 0.012  &  0.022  &  0.308  &  0.334  \\
    Epinions: $+$   & 119,090 & 717,027 &  0.093  &  0.077  &  0.815  &  0.870   \\ \hline
    Slashdot: $-$  &  82,144 & 124,130 & 0.005  &  0.010  &  0.423  &  0.524   \\
    Slashdot: $+$   & 82,144 & 425,072 &  0.025  &  0.022  &  0.906  &  0.909   \\ \hline
    Wikipedia: $-$  & 7,115 & 21,984 &  0.028  &  0.031  &  0.583  &  0.612    \\
    Wikipedia: $+$  & 7,115 & 81,705 &  0.130  &  0.103  &  0.870  &  0.918    \\
  \end{tabular}
 \vspace{-4mm}
  \caption{Networks composed of only positive (negative) edges.
{\em Real:} network induced on the positive (negative) edges.  {\em Rnd:}
network where edge signs are randomly permuted. {\em Clustering:} fraction of
closed triads (closed triads divided by number of length 2 paths) {\em
Component:} fraction of nodes in the largest
connected component. 
} \vspace{-5mm}
  \label{tab:plusMinus}
\end{table}

Table~\ref{tab:plusMinus} summarizes several structural properties of these
networks and their randomized variants. First, we consider the amount of {\em
clustering}, defined as the fraction of $A$-$B$-$C$ paths in which the $A$-$C$
edge is also present (thus forming a ``closed''triad  $A$-$B$-$C$). In all
three datasets, we find that the all-positive networks have significantly
higher clustering than their randomized counterparts, and the all-negative
networks have significantly lower clustering. This further reinforces the
observation that positive edges tend to occur in clumps, while negative edges
tend to span clusters.

Interestingly, both the all-positive and all-negative networks are less
well-connected than expected, in the sense that their largest connected
components are smaller than those of their randomized counterparts. While this
may seem initially counter-intuitive, one possible interpretation is as
follows. The giant components of real social networks are believed to consist
of densely connected clusters linked by less embedded ties
\cite{granovetter-weak-ties,onnela-phone-data}. The all-positive and
all-negative networks in the real (rather than randomized) datasets are each
biased toward one side of this balance: the all-positive networks have dense
clusters without the bridging provided by less embedded ties, while the
all-negative networks lack a sufficient abundance of dense clusters to sustain
a large component.

We also consider the fraction of nodes that are outliers with respect to in-
and out-degree in the all-positive and all-negative networks --- with degrees
exceeding twice the mean for the network. (For reasons of space, these
numerical results are not shown in the table.) These outlier fractions remain
largely unchanged when the edge signs are randomized, with two exceptions that
each hint at interesting conclusions for the effects of displaying signed edges
to users. First, the fraction of outliers for positive in-degree is higher than
expected on Wikipedia, where edge signs are more public. This suggests a
possible tendency for an excess of users to conform to already positive voting
outcomes. Second, the fraction of outliers for negative out-degree is lower
than expected on Epinions and Slashdot, where edge signs are less public. This
is a bit more surprising; it suggests that despite the less public nature of
the signs, there are fewer people who are prolific in their negative
evaluations --- either because the dynamics of these sites suppresses this type
of people, or because they are not attracting people who engage in it.

\section{Conclusion} \label{sec:conclusion}
Social networks underlying current social media sites often reflect a mixture
of positive and negative links. Here we have investigated two theories of
signed social networks --- {\em balance} and {\em status}.  Balance is a
classical theory from social psychology, which in its strongest form postulates
that when considering the relationships between three people, either only one
or all three of the relations should be positive.  Status is a theory of {\em
directed} signed networks which postulates that when person $A$ makes a
positive link to person $B$, then $A$ is asserting that $B$ has higher status
--- with a negative link from $A$ analogously implying that $A$ believes $B$
has lower status. These two theories make different predictions for the
frequency of different patterns of signed links in a social network. On
networks derived from Epinions, Slashdot, and Wikipedia, we find that each
model predicts certain kinds of social relationships, and that there is strong
consistency in how the models fit the data across these three relatively
different settings. Moreover, differences in results between the datasets
highlight some interesting aspects of how the sites present information.

We have discussed the central interpretations of our findings, and here we
briefly review some of the most salient. When the networks are viewed as
undirected graphs, we find strong evidence for a weak form of structural
balance, observing that in all three datasets triangles with exactly two
positive signs are massively underrepresented in the data relative to chance,
while triangles with three positive edges are overrepresented. We further find
that a link is significantly more likely to be positive when its two endpoints
have multiple neighbors (of either sign) in common --- a finding that connects
balance with notions from the theory of social capital. This is particular
pronounced for Wikipedia, where the signs of edges are also the most publicly
prominent.

When the networks are viewed as directed graphs, on the other hand,
incorporating the fact that each link is created by one individual to point to
another, we find that many of the basic predictions of balance theory no longer
apply. Instead, the signs of directed links closely follow the predictions of
the theory of status we develop, in which inferences about the sign of a link
from $A$ to $B$ can be drawn from the mutual relationships that $A$ and $B$
have to third parties $X$.  The signs and directions of these relationships to
$X$ provide information about the status levels of $A$ and $B$, which in turn
accurately predict the deviations in the sign of their interaction from broader
background distributions. Investigating different contexts for links, and the
differences between one-way and reciprocated links, sheds further light on the
subtle ways in which users of these systems draw on behaviors rooted in both
balance and status when they link to one another.

\end{document}